\documentclass[aps,prd,preprintnumbers,nofootinbib,superscriptaddress]{revtex4}
\usepackage{amsmath,amssymb,latexsym}
\usepackage{graphicx}
\usepackage{color}
\newcommand{\X}{\xi}
\newcommand{\thm}{\gamma}
\newcommand{\Tr}{{\rm Tr}}
\newcommand{\tildec}{{\tilde c}}
\newcommand{\HTT}{H}
\newcommand{\hTT}{h}
\newcommand{\calN}{{\cal N}}
\newcommand{\calM}{{\cal M}}
\newcommand{\deltax}{\xi_g}
\newcommand{\deltaxup}{\xi_g}
\newcommand{\rhom}{{\hat\rho_m}}
\newcommand{\rhomg}{{\hat\rho_{m,g}}}
\newcommand{\rhomf}{{\hat\rho_{m,f}}}

\begin{document}
\preprint{IPMU14-0084}
\preprint{YITP-14-23}
\preprint{KUNS-2491}

\title{Viable cosmology in bimetric theory} 

\author{Antonio De Felice}
% \email{}
\affiliation{Yukawa Institute for Theoretical Physics, Kyoto University, Kyoto 606-8502, Japan}

\author{A. Emir G\"umr\"uk\c{c}\"uo\u{g}lu}
% \email{}
\affiliation{School of Mathematical Sciences, University of Nottingham, University Park, Nottingham, NG7 2RD, UK} 
\affiliation{Kavli Institute for the Physics and Mathematics of the Universe, Todai Institutes for Advanced Study, University of Tokyo (WPI), 5-1-5 Kashiwanoha, Kashiwa, Chiba 277-8583, Japan}

\author{Shinji Mukohyama}
% \email{}
\affiliation{Kavli Institute for the Physics and Mathematics of the Universe, Todai Institutes for Advanced Study, University of Tokyo (WPI), 5-1-5 Kashiwanoha, Kashiwa, Chiba 277-8583, Japan}

\author{Norihiro Tanahashi}
% \email{}
\affiliation{Kavli Institute for the Physics and Mathematics of the Universe, Todai Institutes for Advanced Study, University of Tokyo (WPI), 5-1-5 Kashiwanoha, Kashiwa, Chiba 277-8583, Japan}
\affiliation{Department of Applied Mathematics and Theoretical Physics, University of Cambridge, Wilberforce Road, Cambridge CB3 0WA, UK}

\author{Takahiro Tanaka}
% \email{}
\affiliation{Department of Physics, Kyoto University, Kyoto 606-8502, Japan}
\affiliation{Yukawa Institute for Theoretical Physics, Kyoto University, Kyoto 606-8502, Japan}

\date{\today}

\begin{abstract}
We study cosmological perturbations in bimetric theory
with two fluids each of which is coupled to 
one of the two metrics. Focusing on a healthy branch 
of background solutions, we clarify the stability of
the cosmological perturbations.
For this purpose, we extend the condition 
for the absence of the so-called Higuchi ghost, 
and show that the condition is guaranteed to be 
satisfied on the healthy branch. 
We also calculate the squared propagation speeds of 
perturbations and derive the conditions 
for the absence of the gradient instability. 
To avoid the gradient instability, 
we find that the model parameters are weakly constrained. 
\end{abstract}

\maketitle

\section{Introduction}

Our universe consists of various particles with different spins and
masses. The graviton, 
the spin-$2$ particle mediating the gravitational force,
 is of special interest since gravity is the least understood
among fundamental forces in nature. Assuming Lorentz invariance,
Weinberg's theorem in 1964~\cite{Weinberg:1964ew} and its 
extensions~\cite{Weinberg:1980kq,Porrati:2008rm} exclude more than one
interacting massless gravitons in four-dimensional Minkowski
spacetime. Those theorems, however, do not exclude massive graviton(s)
interacting with a massless graviton.

In the present paper we consider a bimetric theory of gravity, {\it
i.e.}~a physical setup involving two dynamical metrics interacting with each
other. In this setup, after diagonalizing the mass matrix for metric
perturbations around Minkowski background, we end up with a massive
graviton and a massless graviton, in accord with the above mentioned
general theorems. Bimetric theory thus propagates seven physical degrees
of freedom in Minkowski background: five from the massive graviton and
two from the massless graviton. Until recently, however, it  was
thought that nonlinear extension of massive gravity inevitably  would have involved a
sixth degree of freedom (eighth degree of freedom in nonlinear bimetric
theory,  {\it e.g.}~\cite{Isham:1971gm}),  which would have been a ghost~\cite{Boulware:1973my}. This ghost
degree of freedom, called Boulware-Deser 
(BD) ghost, was recently 
excised in the
construction of a massive gravity theory by 
de Rham, Gabadadze and Tolley 
(dRGT)~\cite{deRham:2010ik,deRham:2010kj} at fully nonlinear 
level~\cite{Hassan:2011hr,Hassan:2011ea}. A simple extension of the dRGT
massive gravity allows the construction of a fully nonlinear bimetric theory
of gravity without the (would-be) BD ghost~\cite{Hassan:2011zd}. It is
thus this formulation that the studies of bimetric theory in the present
paper are based on.

Having a promising candidate for theoretically consistent bimetric
theory, it is important to investigate whether it can accommodate viable
cosmology. Before starting the study of cosmology in bimetric theory,
however, let us briefly review the current status of cosmology in dRGT
massive gravity and its extensions. 

In the covariant formulation of dRGT massive gravity, the basic
quantities in the gravity sector are a metric field and four scalar 
fields called St\"{u}ckelberg fields. The original dRGT theory respects
the Poincar\'e symmetry in the space of St\"{u}ckelberg fields so that
the St\"{u}ckelberg fields enter the action only through the so-called
fiducial metric, which is the pullback of Minkowski metric in the field
space to the spacetime. It turned out that the dRGT theory with the
internal Poincar\'e symmetry does not allow for any non-trivial flat
Friedmann-Lema\^\i tre-Robertson-Walker (FLRW)
solutions~\cite{D'Amico:2011jj}. The same no-go result holds for closed
FLRW solutions as well. One can nonetheless find non-trivial and
self-accelerating open FLRW solutions in this 
theory~\cite{Gumrukcuoglu:2011ew}. Slightly extending the theory by 
replacing the Minkowski fiducial metric with de Sitter or FLRW one, it
also becomes possible to find not only open but also flat and closed
FLRW solutions with~\cite{Gumrukcuoglu:2011zh} or
without~\cite{Hassan:2011vm,Fasiello:2012rw,Langlois:2012hk} 
self-acceleration. Unfortunately, all those FLRW solutions in dRGT
theory turned out to be unstable due to either linear instability called
Higuchi ghost~\cite{Higuchi:1986py} or a recently found nonlinear
instability~\cite{DeFelice:2012mx,DeFelice:2013awa}.

The origin of the new nonlinear instability found
in~\cite{DeFelice:2012mx} is the fact that kinetic terms of three among
five degrees of freedom in FLRW backgrounds are exactly proportional to
the equation of motion for the temporal St\"{u}ckelberg field and thus
vanish on shell~\cite{Gumrukcuoglu:2011zh}. Those kinetic terms vanish
at the quadratic level in the action, but do appear at the nonlinear level
and can become either positive or negative, depending on the nature of
perturbations.

One can in principle evade the instability of cosmological solutions in
dRGT massive gravity by relaxing the FLRW symmetry, {\it i.e.}~either
homogeneity~\cite{D'Amico:2011jj}  (see
also~\cite{Chamseddine:2011bu,Koyama:2011xz,Koyama:2011yg,Gratia:2012wt,Kobayashi:2012fz,Volkov:2012cf}
for related solutions) or
isotropy~\cite{Gumrukcuoglu:2012aa,DeFelice:2013awa}. Another 
possibility is to extend the theory by introducing an extra scalar field
in the gravity
sector~\cite{D'Amico:2012zv,Huang:2012pe,DeFelice:2013tsa}. In both
cases, the above mentioned exact proportionality between the kinetic
terms and the background equation of motion is detuned and thus the
nonlinear instability can in principle be avoided. 

The nonlinear bimetric theory of gravity can be considered as yet
another extension of dRGT massive gravity, due to an extra spin-$2$
field. Hence, the above mentioned no-go result for stable FLRW
cosmological solution does not directly apply to the nonlinear bimetric
theory. Unfortunately, in nonlinear bimetric theory 
the analogue of the self-accelerating branch 
still suffers from nonlinear instability. We thus study
the stability of FLRW solutions in the other branch in
the bimetric theory  originally proposed 
in the context of the graviton oscillations and gravitational wave
observations \cite{DeFelice:2013nba}, 
which we call the healthy branch.~\footnote{Similar cosmological background solutions in bimetric theory with a single matter source have been considered in
\cite{Volkov:2011an, Comelli:2011zm}, while Ref.~\cite{Maeda:2013bha} studied backgrounds with two fluids.
}
In this study, we take into account two matter fields each
of which couples to either the first or second metric. 

The rest of the present paper is organized as follows.
In Sec.~\ref{sec:model} we describe the model of our interest. 
We briefly review the background cosmology assuming 
the spatial homogeneity and isotropy in Sec.~\ref{sec:background}, 
where we also introduce the notion of two different branches. 
Then, we discuss linear perturbations around this cosmological 
background in Sec.~\ref{sec:perturbation}. Starting with the pure
gravity case, we discuss the tensor, vector and scalar-type
perturbations one by one. We identify the conditions for 
the absence of ghost and gradient instability.
Section~\ref{sec:summary} is devoted to summary and discussion. 

\section{Model}
\label{sec:model}

The covariant action for the gravity sector is constructed out of two
four-dimensional metrics $g_{\mu\nu}$ and $f_{\mu\nu}$. It is the sum of
the two Einstein-Hilbert actions $I_{EH,g}$ and $I_{{\rm EH},f}$ for the metrics
$g_{\mu\nu}$ and $f_{\mu\nu}$, respectively, and 
the non-derivative mixing term $I_{\rm mix}$ 
which is built by requiring that the Boulware--Deser ghost is absent at all orders~\cite{Hassan:2011hr}.
Including the matter, the total action we consider is 
\begin{equation}
 I = I_{{\rm EH},g} + I_{{\rm EH},f} + I_{\rm mix} + I_{\rm matter},
\end{equation}
where 
\begin{eqnarray}
 I_{{\rm EH},g} &\! =\! & 
  \frac{M_g^2}{2}\int d^4x\sqrt{-g}R[g], 
  \\
 I_{{\rm EH},f} &\! =\! & 
  \frac{M_f^2}{2}\int d^4x\sqrt{-f}R[f],
  \\
 I_{\rm mix} &\! =\! & 
  m^2 M_g^2\int d^4x\sqrt{-g}\,
\sum_{i=0}^4 \alpha_i {\cal L}_i\,.
\label{Imix}
\end{eqnarray}
Each term in $I_{\rm mix}$ is constructed as 
\begin{eqnarray}
{\cal L}_0 &\! =\! & 
1\,, \qquad
{\cal L}_1 =  
\left[{\cal K}\right]\,,
\,\qquad 
 {\cal L}_2 = \frac{1}{2}
  \left(\left[{\cal K}\right]^2-\left[{\cal K}^2\right]\right)\,, \qquad
 {\cal L}_3 = \frac{1}{6}
  \left(\left[{\cal K}\right]^3-3\left[{\cal K}\right]\left[{\cal K}^2\right]+2\left[{\cal K}^3\right]\right), 
  \qquad
\cr
 {\cal L}_4 &\! =\! & \frac{1}{24}
  \left(\left[{\cal K}\right]^4-6\left[{\cal K}\right]^2\left[{\cal K}^2\right]+3\left[{\cal K}^2\right]^2
   +8\left[{\cal K}\right]\left[{\cal K}^3\right]-6\left[{\cal K}^4\right]\right)\,,
\end{eqnarray}
where the square brackets denote trace operation and 
\begin{equation}
{\cal K}^\mu _{\,\nu} = \delta^\mu_{\,\nu} 
 - \left(\sqrt{g^{-1}f}\right)
\!\raisebox{0.5mm}{${}^{\mu}$}\raisebox{-0.5mm}{$\!_{\nu}$}
\,.
\end{equation}
The square root in this expression represents the 
matrix that satisfies 
\begin{equation}
 \left(\sqrt{g^{-1}f}\right)
\!\raisebox{0.5mm}{${}^{\mu}$}\raisebox{-0.5mm}{$\!_{\rho}$}
  \left(\sqrt{g^{-1}f}\right)
\!\raisebox{0.5mm}{${}^{\rho}$}\raisebox{-0.5mm}{$\!_{\nu}$}
  = g^{\mu\rho}f_{\rho\nu}\,, 
\end{equation}
and whose eigenvalues are all positive. 
Notice that the cosmological constant term for $f_{\mu\nu}$ can be
expressed as a linear combination of ${\cal L}_i$ ($i=0,\ldots,4$),
while that for $g_{\mu\nu}$ is ${\cal L}_0$. Besides the three mass scales
$M_g$, $M_f$ and $\sqrt{{\rm Max}_i|\alpha_i|}m$, we thus have four
dimensionless parameters (four among $\alpha_i$ ($i=0,\ldots,4$)).
This means that, for each choice of the set of mass scales $M_g$,
$M_f$ and $\sqrt{{\rm Max}_i|\alpha_i|}m$, we have freedom to tune
four additional quantities, {\it e.g.} $\X_c$, $J(\X_c)$, $J'(\X_c)$ and
$\Lambda/m^2$ defined in the next section.

As for the matter sector, we consider scalar fields $\phi_g$ and $\phi_f$
minimally coupled to $g_{\mu\nu}$ and $f_{\mu\nu}$, respectively. Because of their equivalence to irrotational barotropic perfect fluids (see {\it e.g.}~\cite{kessence-perfectfluid}), we restrict our consideration to k-essence fields\footnote{The effect of rotational perturbations are briefly discussed in the Appendix.}, with the action
\begin{equation}
 I_{\rm matter} = \int\, d^4x\sqrt{-g}P_g(X_g)
  + \int\, d^4x\sqrt{-f}P_f(X_f)\,,
\end{equation}
with
\begin{eqnarray}
  X_g \equiv -g^{\mu\nu}\partial_{\mu}\phi_g\partial_{\nu}\phi_g\,,
  \qquad
  X_f \equiv -f^{\mu\nu}\partial_{\mu}\phi_f\partial_{\nu}\phi_f\,. 
\label{matter}
\end{eqnarray}
When we simply write $P_a$, with $a=g,f$, it denotes 
the background value of $P_a(X_a)$ and corresponds to the background 
pressure in the perfect fluid picture. We also introduce the energy
density and
sound speed of the effective fluid as 
\begin{equation}
 \rho_a \equiv 2P_a'(X_a)X_a - P_a(X_a), \qquad
 c_a^2 \equiv \frac{P_a'(X_a)}{2P_a''(X_a)X_a+P_a'(X_a)}\,,
\end{equation}  
where prime denotes derivative with respect to the argument. 

\section{Background equations and Solution branches}
\label{sec:background}
In this section, 
we derive the equations of motion for the background FLRW universe,
and also discuss the branches of the background solutions. 
We denote the two metrics $g_{\mu\nu}$ and $f_{\mu\nu}$ as 
\begin{eqnarray}
&& g_{\mu\nu}dx^{\mu}dx^{\nu} 
  = -N^2 dt^2
  + a^2 \thm_{ij} dx^idx^j\,, \nonumber\\
&& f_{\mu\nu}dx^{\mu}dx^{\nu} 
  = -\ n^2 dt^2
  + \alpha^2 \thm_{ij} dx^idx^j\,,
\end{eqnarray}
with
\begin{eqnarray}
 \thm_{ij} & \equiv & \delta_{ij} + 
  \frac{K\delta_{il}\delta_{jm}x^lx^m}{1-K\delta_{lm}x^lx^m}\,,
\end{eqnarray}
where $N=N(t)$ and $n=n(t)$ are 
the background lapse functions, and 
$a=a(t)$ and
$\alpha=\alpha(t)$ are background scale factors.
Notice that the two background metrics are 
both homogeneous and isotropic, having common isometries. 
The background equations are given by 
\begin{eqnarray}
&&3\left(H^2 + \frac{K}{a^2}\right) = %\Lambda_g+
m^2\,\rhomg +\frac{\rho_g}{M_g^2}\,,
\label{eqn}\\
&&3\left(H_f^2 + \frac{K}{\alpha^2}\right) =
\frac{m^2}{\kappa}\,
\rhomf +
\frac{\rho_f}{\kappa\, M_g^2}
\,,
\label{eqnf}\\
&&
2 \left( \frac{\dot{H}}{N}-\frac{K}{a^2}\right)= m^2 \X J (\tildec-1) - \frac{\rho_g + P_g}{M_g^2} \,,
\label{eqh}\\
&&
2\left(\frac{\dot{H}_f}{n}-\frac{K}{\alpha^2}\right)= 
-\frac{m^2}{\kappa\, \X^3 \tildec}\,
J (\tildec-1)- 
\frac{\rho_f+P_f}{\kappa\, M_g^2}
\,,
\label{eqhf}\\
&&\left[\frac{1}{c_g^2}\frac{1}{N}\frac{d}{dt}
   \left(\frac{\dot{\phi}_{g,0}}{N}\right) + \frac{3 H \dot{\phi}_{g,0}}{N}\right](\rho_g+P_g)=0\,,
\label{eqc}\\
&&\left[\frac{1}{c_f^2}\frac{1}{n}\frac{d}{dt}\,\left(\frac{\dot{\phi}_{f,0}}{n}\right)
   + \frac{3 H_f \dot{\phi}_{f,0}}{n}\right](\rho_f+P_f)=0\,,\label{eqt}
\end{eqnarray}
where we have defined $H\equiv\dot{a}/(Na)$, 
$H_f\equiv \dot{\alpha}/(n\alpha)$, $\kappa \equiv M_f^2 / M_g^2$ and an
overdot represents derivative with respect to $t$. 
Here, we have also introduced
\begin{eqnarray}
 \rhomg  &\! 
\equiv
\! & 
  U(\X)-{\X\over 4}U'(\X)
\,,
  \qquad
 \rhomf  
\equiv
{1 \over 4\X^3} U'(\X)
\,,
  \nonumber\\
 J(\X) & \equiv & 
{1 \over 3}\left[U(\X)-{\X\over 4}U'(\X)\right]' 
\,,
\label{bgdefs}
\end{eqnarray}
where $\X$ and $\tildec$ are, respectively, 
the ratios of the background scale factors 
and lapse functions defined as
\begin{equation}
 \X  \equiv  \frac{\alpha}{a}\,, 
\qquad 
\tildec \equiv \frac{n a}{N \alpha}\,, 
\end{equation}
and 
\begin{equation}
 U(\X)\equiv -\alpha_0+4(\X-1)\alpha_1-6(\X-1)^2\alpha_2+4(\X-1)^3\alpha_3-
(\X-1)^4\alpha_4\,.
\end{equation}
For later convenience, we define 
\begin{eqnarray}
   \Gamma(\X) &\equiv &\X J(\X) + \frac{(\tildec -1) \X^2}{2} J'(\X)
  \,, \label{Gamma} \\
  m_{\rm eff}^2(\X)  &\equiv &
   \frac{1 + \kappa \X^2}{\kappa \X^2}m^2\Gamma(\X)\,.
   \label{meff}
\end{eqnarray}
In the next section we shall see that $m_{\rm eff}$ is the effective
graviton mass.~\footnote{
We remark that when the light cones of the two metrics coincide, {\it i.e.} $\tilde{c}=1$, the effective graviton mass reduces to the Fierz-Pauli mass given in \cite{Hassan:2012wr}, with the following correspondence between the two notations:
\begin{eqnarray*}
&&\alpha_0 \to -\beta_0 - 4 \beta_1 - 6 \beta_2 - 
   4 \beta_3 - \beta_4
\,,\qquad
 \alpha_1 \to \beta_1 + 3 \beta_2 + 
   3 \beta_3 + \beta_4
\,,\qquad
 \alpha_2 \to -\beta_2 - 
   2 \beta_3 - \beta_4
\,,\qquad \alpha_3 \to \beta_3 + \beta_4
\,,\qquad \alpha_4
\to -\beta_4\,,\nonumber\\
&&
\qquad\qquad\qquad\qquad\qquad\qquad\qquad\qquad
M_g\to \sqrt{2}m_g\,,\qquad
M_f\to \sqrt{2}m_f\,,\qquad
m\to m^2/m_g\,.
\end{eqnarray*}
}

Combining Eqs.~(\ref{eqn}), (\ref{eqh}) and (\ref{eqc}), or
equivalently, Eqs.~(\ref{eqnf}), (\ref{eqhf}) and (\ref{eqt}), 
we obtain a constraint, 
\begin{equation}
J(H-\X H_f) = 0\,.
\end{equation}
This constraint indicates that there are two branches of solutions: one
is specified by the condition $J=0$ and the other by $H = \X H_f$.

 Before starting the analysis of the healthy branch,
we briefly mention the  other branch with $J=0$.
Since the condition $J=0$ implies that $\X=\alpha/a$ is constant, the quantities
$\rhomg $ and $\rhomf$ are also constant. 
In this case the quadratic order of perturbation of the 
mixing term of the action, $I_{\rm mix}$, 
is equivalent to the cosmological constant terms 
on both metrics. (See Eq.~\eqref{mixquad} below with $J$ 
being set to 0.) 
Thus, the Hamiltonian structure for linear perturbations is 
the same as that in 
two copies of general relativity. As a result, instead of the $7$
degrees of freedom we expect from a healthy bimetric theory, we end up
with $4$ degrees of freedom that are dynamical at linear
order.~\cite{Comelli:2012db}. This branch is an analogue of the
self-accelerating branch in massive gravity~\cite{Gumrukcuoglu:2011ew},
which is known to possess incurable problems, 
such as  a less-than-expected number of the degrees of freedom at quadratic level~\cite{Gumrukcuoglu:2011zh} and appearance of a non-perturbative ghost~\cite{DeFelice:2012mx}.
Thus, we consider only the healthy branch specified by $H=\X H_f$ in the
rest of this paper.

First, we derive two algebraic relations, {\it i.e.}~constraints that
hold in 
the $H = \X H_f$ branch.
Combining Eqs.~(\ref{eqn}) and (\ref{eqnf}) under the assumption 
$H = \X H_f$, we have 
\begin{equation} 
m^2 \rhom (\X)
=-
\frac{\rho_g}{M_g^2} 
+
\frac{\X^2 \rho_f}{\kappa M_g^2}
\,, 
\label{Fconst1}
\end{equation}
where we have introduced a function of $\X$ defined by 
\begin{equation}
\rhom (\X)
\equiv \rhomg (\X)-{\X^2 \over \kappa}\rhomf (\X)
      =U(\X)-{1\over 4}\left({\X+{1\over \kappa \X}}\right)U'(\X)\,.
\end{equation}
Equation~(\ref{Fconst1}) should be interpreted as the equation that determines 
the value of $\X$. 
The other constraint is derived from 
$d(H - \X H_f)/dt=0$.
Using Eqs.~(\ref{eqh}) and (\ref{eqhf}), this constraint can be rewritten as
\begin{equation}
2 (\tildec -1) W
=\frac{\rho_g + P_g}{M_g^2}
 -\frac{\tildec \X^2 (\rho_f + P_f)}{\kappa M_g^2}\,, 
\label{Fconst2}
\end{equation}
where we have defined 
\begin{equation}
W\equiv \frac{m^2 (1+\kappa \X^2)J}{2\kappa \X}-H^2-\frac{K}{a^2}\,.
\end{equation}
This constraint is to be 
interpreted as the equation that determines the difference  
between the light cones of two metrics, $\tildec -1$. 
Hence, 
barring
special tuning of model parameters, vanishing of 
$(\rho_g + P_g) - \X^2 (\rho_f + P_f)/\kappa$ 
means the crossing 
of $\tilde c=1$ and $W$ keeps a definite sign along the
trajectory of the time evolution. 
The regime $\tildec > 1$ 
is preferred as 
a viable model when the ordinary matter fields are coupled to the 
$g$ metric. This is because, when $\tildec <1$, the electro-magnetic
wave travels faster than the propagation speed of the $f$-metric
perturbations and hence 
a UHECR traveling with a speed very close the speed of light may emit the 
$f$-gravi-Cherenkov radiation, which is severely constrained by
observations~\cite{Moore:2001bv,Kimura:2011qn}.

In Ref.~\cite{DeFelice:2013nba} 
a healthy background cosmology was proposed. 
We generalize the analysis in \cite{DeFelice:2013nba} to the case with
two matter fields, removing also the restriction to the low energy regime. 
In this solution at low energies, 
\begin{equation}
\frac{\rho_g}{m^2M_g^2} \ll 1\,, \quad
 \frac{\xi^2\rho_f}{\kappa m^2M_g^2} \ll 1\,,
\label{lowenergy}
\end{equation}
$\X$ converges to a constant $\X_c$ (assumed to be $\X_c =O(1)$) that solves 
\begin{equation}
  \rhom (\X_c)=0\,.  
\end{equation}
The mass scale of the coupling term $m$ is assumed to be 
parametrically large, but the effective graviton mass 
$m_{\rm eff}(\X_c)$ is kept small, 
by tuning the model parameters. 

The other constraint~(\ref{Fconst2}) implies $\tildec \to 1$ 
at low energies. Namely, the difference of the light cone between 
two metrics vanishes in the low energy limit. 
We expand $\rhom (\X)$ around $\X=\X_c$ as 
\begin{eqnarray} 
\rhom (\X)
&\simeq&
\left.\frac{d\rhom (\X)}{d\X}\right\vert_{\X=\X_c} \!\!\! (\X-\X_c)
= \left[
   \frac{3(1+\kappa \X_c^2)J_c}{\kappa \X_c^2}
   - \frac{2\Lambda}{\X_c m^2}\right](\X-\X_c)\,,
\end{eqnarray}
where we defined 
\begin{equation}
 J_c\equiv J(\X_c), \qquad
  \Lambda \equiv m^2\rhomg (\X_c)\, . 
\label{effectivelambda}
\end{equation}
Hence, Eq.~(\ref{Fconst1}) implies that
\begin{equation}
 m^2
  \left[
   \frac{3(1+\kappa \X_c^2)J_c}{\kappa \X_c^2}
   - \frac{2\Lambda}{\X_c m^2}
        \right](\X-\X_c) \simeq 
  -\frac{\rho_g}{M_g^2}+\frac{\X_c^2\rho_f}{\kappa M_g^2}
  \,.
\end{equation}
Using this, we find that the Friedmann equation~(\ref{eqn}) can be
approximated as 
\begin{equation}
3\left( H^2 + \frac{K}{a^2}\right)
=
\frac{\rho_g}{M_g^2} + m^2 \rhomg (\X)
\simeq
\frac{\rho_g}{M_g^2} + m^2 \rhomg (\X_c) + 3 m^2 J_c (\X-\X_c)
\simeq
\frac{\rho_g+\tilde{\kappa}^{-1}\rho_f}{\tilde{M}_g^2}
+ \Lambda\,,
\label{effFeq}
\end{equation}
where we defined 
\begin{equation}
 \tilde{M}_g^2 \equiv 
\left[1 + 
\frac{3\kappa\X_c^2m^2J_c}{3m^2J_c-2\kappa\X_c\Lambda}
\right]
M_g^2, \quad
 \tilde{\kappa} \equiv 
 \frac{1}{\X_c^4}
 - \frac{2\kappa\Lambda}{3\X_c^3m^2J_c}. 
\end{equation}
Eq.~(\ref{effFeq}) can be interpreted as the Friedmann equation for two
matter fields $\rho_g$ and $\tilde{\kappa}^{-1}\rho_f$ with the
effective gravitational constant $\tilde{M}_g^2$ and the effective
cosmological constant $\Lambda$. We did not assume smallness 
of $|\Lambda/m^2|$ so far. 
The pure gravity case can be easily obtained by taking the 
limit of $\rho_a=0$ and $P_a=0$. In this case both metrics 
are de Sitter and we have $\X=\X_c$ and $\tildec =1$.

If we tune the effective cosmological constant $\Lambda$ so that  
\begin{equation}
\left|\frac{\kappa\X_c\Lambda}{m^2J_c}\right| \ll 1\,,
\label{tuningalpha0}
\end{equation}
then 
\begin{equation}
 \tilde{M}_g^2 \simeq M_+^2 \equiv 
\left({1 + \kappa \X_c^2}\right) M_g^2, \qquad
\tilde{\kappa} \simeq \frac{1}{\X_c^4}\,.
\label{effMg}
\end{equation}
Hereafter, we assume Eq.~\eqref{tuningalpha0} as well as Eq.~(\ref{lowenergy}) to hold when we take the low energy limit.
We remark that we do not intend to solve the cosmological constant
problem in the present paper and that the condition (\ref{tuningalpha0})
can be realized by simply tuning the $\alpha_0$ parameter.

In the low energy limit or in the pure gravity case, 
$W$ in Eq.~\eqref{Fconst2} 
with $K=0$ is reduced to 
$m_{\rm eff}^2/2-H^2$, where $m_{\rm eff}$ is the effective graviton
mass defined in Eq.~(\ref{meff}).
This quantity 
must be positive for the absence of Higuchi
ghost~\cite{Higuchi:1986py}. 
As mentioned above, the sign of $W$ does not change in general. 
Therefore, we choose the branch in which 
\begin{equation}
W >0
\label{branchcondiJ}
\end{equation}
is satisfied. In this case, the condition to avoid 
the Cherenkov radiation, $\tildec > 1$, is 
reduced to 
\begin{equation}
 \rho_g + P_g > \frac{\tildec\, \X^2}{\kappa} (\rho_f + P_f)\,.
\end{equation}
The positivity of $W$ also indicates that $J=0$ is not 
realized except 
in the limit where $H^2+K/a^2$ and $W$ simultaneously vanish, provided
that $|K|/a^2<H^2$ in accord with observation. 
However, 
in this low energy limit, the value of $\X$ converges to $\X_c$, 
which is different from the zeros of $J(\X)$ in general. 
In other words, $J(\X_c)\ne 0$ unless fine-tuned. 
Hence, we find that the healthy branch does not cross the $J=0$ branch.

The positivity of $W$ may break down only when 
the sequence of solution disappears as we increase the energy scale. 
From the constraint~\eqref{Fconst2}, we find 
that $\tildec -1$ diverges
when $W$ crosses 0. 
Notice that we can also express $\tildec -1$ as 
$$
 \tildec -1 = {\dot \X\over N H\X}\,. 
$$
Hence, when $\tildec -1$ diverges, $\dot \X$ also diverges. 
This indicates that 
the system would exit the regime of validity of the effective field
theory there. 
By contrast, when the right hand side of Eq.~\eqref{Fconst2}
vanishes, it just means $\tildec -1$ crosses zero in general. 
Therefore, this does not mean the flip of the sign of $W$. Hence, we 
conclude that, 
in the regime of validity of the effective field theory, 
the condition $W>0$ is maintained.

The above phenomena can be understood more intuitively when 
there is no matter field coupled to the $f$-metric. 
Notice that $W$ is related to $d\rhom /d\ln\X$ as 
\begin{equation}
 {d\rhom \over d\ln\X}
   =3\left(\X+{1\over\kappa\X}\right)J -{2\X^2\rhomf \over \kappa}
   ={6W\over m^2}+{2\X^2 \rho_{f}\over \kappa m^2 M_g^2}\, . 
\end{equation}
Hence, when there is no matter coupled to the $f$-metric, 
$W>0$ is identical to ${d\rhom/d\ln \X}>0$, 
which is the condition for the absence of Higuchi ghost discussed 
in Ref.~\cite{Yamashita:2014cra}. 
The meaning of this condition is clear: as we increase the matter energy density $\rho_g$, the
constraint~\eqref{Fconst1} with $\rho_f=0$ implies that 
$\rhom $ should decrease. When the minimum of the function 
$\rhom (\X)$ is reached, 
we cannot extend the background solution beyond that critical 
energy density of $\rho_g$. 
Conversely, as long as the healthy branch solution continues to exist,  
the condition ${d \rhom/d\ln \X}>0$ is maintained.

\section{Perturbations around FLRW backgrounds}
\label{sec:perturbation}
\subsection{Pure gravity case}
We expand the two metrics perturbed around FLRW
backgrounds as
\begin{eqnarray}
&& g_{\mu\nu}dx^{\mu}dx^{\nu} 
  = -N^2 (1+2\Phi)\,dt^2
  + 2N a V_i \,dt\,dx^i
  + a^2(\thm_{ij}+H_{ij}) dx^idx^j\,, \nonumber\\
&& f_{\mu\nu}dx^{\mu}dx^{\nu} 
  = -\ n^2 (1+2 \varphi)\, dt^2
  + 2 n \alpha v_i\, dt\,dx^i
  + \alpha^2(\thm_{ij}+h_{ij}) dx^idx^j\,, 
\end{eqnarray}
where 
($\Phi$, $V_i$,
$H_{ij}$) and ($\varphi$, $v_i$, $h_{ij}$) represent perturbations. 
For the present discussion, it is convenient to decompose $H_{ij}$ and
$h_{ij}$ further into their trace parts and traceless parts as
\begin{equation}
 H_{ij} = \frac{1}{3}\thm_{ij}\Tr H  + H^T_{ij}\,,\quad
 h_{ij} = \frac{1}{3}\thm_{ij}\Tr h + h^T_{ij}\,,
\end{equation}
where $\Tr H\equiv\thm^{ij}H_{ij}$, 
$H^T_{ij}\equiv H_{ij}-\frac{1}{3}\thm_{ij}\Tr H$, 
$\Tr h\equiv\thm^{ij}h_{ij}$, and 
$h^T_{ij}\equiv h_{ij}-\frac{1}{3}\thm_{ij}\Tr h$. 

The Einstein-Hilbert action for $g_{\mu\nu}$ is expanded as 
\begin{equation}
 I_{{\rm EH},g} = \frac{M_g^2}{2}
  \int d^4x\, N a^3 \sqrt{\thm}
  \left( L_{EH,g}^{(0)} + L_{EH,g}^{(1)} + L_{EH,g}^{(2)} + \cdots
  \right)\,,
\end{equation} 
and the 
quadratic part is given by 
\begin{eqnarray}
 L_{EH,g}^{(2)} &\! =\! & 
  \frac{1}{4N^2}
  \left(\thm^{ik}\thm^{jl}
   -\thm^{ij}\thm^{kl}\right)\dot{H}_{ij}\dot{H}_{kl}
  + \frac{2H}{N}\Phi \Tr \dot{H}
  \nonumber\\
 & & 
  + \frac{1}{Na}\thm^{ij}V_iD^{k}\dot{H}^T_{jk}
  + \frac{2}{3Na}\Tr \dot{H}D^kV_k
  - \frac{4H}{a}\Phi D^kV_k \nonumber\\
 & & 
 + \frac{2}{3a^2}D^i\Phi D_i\Tr H
 + \frac{1}{18a^2}D^i\Tr HD_i\Tr H
 - \frac{1}{a^2}D^i\Phi D^jH^T_{ij} \nonumber\\
 & & 
 - \frac{1}{6a^2}D^i\Tr HD^jH^T_{ij}
 + \frac{1}{2a^2}\thm^{ij}D^kH^T_{ik}D^lH^T_{jl}
 - \frac{1}{4a^2}\thm^{jl}\thm^{km}D^iH^T_{jk}D_iH^T_{lm}
 \nonumber\\
 & & 
  + \frac{1}{4a^2}\thm^{ik}\thm^{jl}F_{ij}F_{kl}
  + \left(-9H^2-\frac{3K}{a^2}
\right)\Phi^2
  + \left(3H^2+\frac{K}{a^2}
\right)
  (\Phi \Tr H+\thm^{ij}V_iV_j) \nonumber\\
 & & 
  + \left(\frac{1}{4}H^2+\frac{1}{6}\frac{\dot{H}}{N}
     -\frac{K}{12a^2}
\right)
  (\Tr H)^2
  + \left(-\frac{3}{2}H^2-\frac{\dot{H}}{N}-\frac{K}{a^2}
\right)
  \thm^{ik}\thm^{jl}H^T_{ij}H^T_{kl}\,,
\end{eqnarray}
where $F_{ij}\equiv\partial_iV_j-\partial_jV_i$, 
$D_i$ is the covariant differentiation with respect to 
$\thm_{ij}$, and $D^i\equiv \thm^{ij}D_j$. 
A similar expansion applies to the Einstein-Hilbert action for
the $f$-metric.

Up to second order, 
the mixing term~(\ref{Imix}) is expanded as 
\begin{eqnarray}
 \frac{I_{\rm mix}}{m^2 M_g^2} &\! =\! & 
  \int d^4x\, 
  \left[ -\sqrt{-g}\, \rhomg - \sqrt{-f}\, \rhomf
   + \frac{1}{2}N a^3 \sqrt{\thm}\, \X J \Delta
  \right] \nonumber\\
 & &  + \frac{1}{8}
  \int d^4x\, N a^3 \sqrt{\thm}\, \Gamma 
  (\thm^{ij}\thm^{kl}-\thm^{ik}\thm^{jl})
  (H_{ij}-h_{ij})(H_{kl}-h_{kl})\,,
\label{mixquad}
\end{eqnarray}
with 
\begin{eqnarray}
 \frac{\sqrt{-g}}{N a^3 \sqrt{\thm}} &\! =\! & 
  1 + \left(\Phi+\frac{1}{2}\Tr H\right)
  + \left[
     -\frac{1}{2}\Phi^2 + \frac{1}{2}\thm^{ij}V_iV_j
     +\frac{1}{8}\left(\thm^{ij}\thm^{kl}
     - 2\thm^{ik}\thm^{jl}\right)H_{ij}H_{kl}
     + \frac{1}{2}\Phi \Tr H \right] 
\,, \nonumber\\
 \frac{\sqrt{-f}}{n \alpha^3 \sqrt{\thm}} &\! =\! & 
  1 + \left(\phi+\frac{1}{2}\Tr h\right)
  + \left[
     -\frac{1}{2}\varphi^2 + \frac{1}{2}\thm^{ij}v_iv_j
     +\frac{1}{8}\left(\thm^{ij}\thm^{kl}
     - 2\thm^{ik}\thm^{jl}\right)h_{ij}h_{kl}
     + \frac{1}{2}\varphi \Tr h \right] 
\,, \nonumber\\
 \Delta &\! =\! & -(\tildec-1)(\Tr H-\Tr h)
  + (\Phi-\tildec\varphi) (\Tr H-\Tr h)
  + \frac{1}{\tilde{c}+1}\thm^{ij}(V-\tildec v)_i(V-\tildec v)_j
  \nonumber\\
 & & - \frac{\tildec-1}{4}
  \left(\thm^{ij}\thm^{kl}-2\thm^{ik}\thm^{jl}
       \right)(H+h)_{ij}(H-h)_{kl}\,.
\end{eqnarray} 

In the pure gravity limit, 
the quadratic action can then be diagonalized by
introducing new perturbation variables 
\begin{eqnarray}
&& \Phi^- \equiv \Phi-\varphi\,, \quad
  V^-_i \equiv V_i-v_i\,, \quad
  H^-_{ij} \equiv H_{ij}-h_{ij}\,,\cr
&& \Phi^+ \equiv 
\frac{\Phi+\kappa \X^2 \varphi}{1+\kappa \X^2}\,,
\quad
  V^+_i \equiv 
\frac{V_i+\kappa \X^2 v_i}{1+\kappa \X^2}\,, 
\quad
  H^+_{ij} \equiv 
\frac{H_{ij}+\kappa \X^2 h_{ij}}{1+\kappa \X^2}\,. 
\label{fieldredef}
\end{eqnarray}
Using the relations for the background, 
we can rewrite the quadratic action for perturbation as 
\begin{eqnarray}
 I^{(2)} &\! =\! & \frac{1}{2}\int d^4x\, N a^3 \sqrt{\thm} L^{(2)}\,,
\end{eqnarray}
with
\begin{eqnarray}
 L^{(2)} = M_+^2 L^{(2)}_{\rm EH}[\Phi^+,V^+_i,H^+_{ij}; \Lambda]
  + M_-^2 \left(L^{(2)}_{\rm EH}[\Phi^-,V^-_i,H^-_{ij}; \Lambda]
 + m_{\rm eff}^2 L^{(2)}_{\rm FP}[\Phi^-,V^-_i,H^-_{ij}]
	\right)\, ,
\label{Lpuregrav}
\end{eqnarray} 
where 
\begin{eqnarray}
 L^{(2)}_{\rm EH}[\Phi,V_i,H_{ij}; \Lambda] &\! =\! & 
  \frac{1}{4N^2}
  \left(\thm^{ik}\thm^{jl}
   -\thm^{ij}\thm^{kl}\right)\dot{H}_{ij}\dot{H}_{kl}
  + \frac{2H}{N}\Phi \Tr \dot{H}
  + \frac{1}{Na}\thm^{ij}V_iD^{k}\dot{H}^T_{jk}
  \nonumber\\
 & & 
  + \frac{2}{3Na}\Tr \dot{H}D^kV_k
  - \frac{4H}{a}\Phi D^kV_k 
 + \frac{2}{3a^2}D^i\Phi D_i\Tr H 
 - \frac{1}{a^2}D^i\Phi D^jH^T_{ij} 
 \nonumber\\
 & & 
 + \frac{1}{18a^2}D^i\Tr HD_i\Tr H
 - \frac{1}{6a^2}D^i\Tr HD^jH^T_{ij}
 + \frac{1}{2a^2}\thm^{ij}D^kH^T_{ik}D^lH^T_{jl}
\nonumber\\
 & & 
 - \frac{1}{4a^2}\thm^{jl}\thm^{km}D^iH^T_{jk}D_iH^T_{lm}
  + \frac{1}{4a^2}\thm^{ik}\thm^{jl}F_{ij}F_{kl}
  - 2\Lambda\Phi^2 
\nonumber\\
 & & 
+\frac{K}{a^2}\left[6\Phi^2-2(\Phi \Tr H+\thm^{ij} V_iV_j)-\frac{1}{6}(\Tr  H)^2-\frac{1}{2}\thm^{ik}\thm^{jl}H_{ij}^TH_{kl}^T\right],
\nonumber\\
 L^{(2)}_{\rm FP}[\Phi^-,V^-_i,H^-_{ij}] &\! =\! & 
  \Phi^-\Tr H^-+\frac{1}{2}\thm^{ij}V^-_iV^-_j
  + \frac{1}{4}\left(\thm^{ij}\thm^{kl}
		-\thm^{ik}\thm^{jl}\right)H^-_{ij}H^-_{kl}\,.
\end{eqnarray}
The effective gravitational coupling for the ``$+$'' fields, $M_+^2$,  
the effective cosmological constant, $\Lambda$, and the effective
graviton mass for the ``$-$'' fields, $m_{\rm eff}$, have been already
defined in Eqs.~\eqref{effMg}, \eqref{effectivelambda}\,\footnote{
Nevertheless, $|\Lambda/m^2|$ does not have to be small in this
section. The background equation is given by (\ref{effFeq})  with
$\rho_g=\rho_f=0$.} 
and \eqref{meff}, 
while the effective gravitational coupling for the ``$-$'' fields is given by
\begin{equation}
\qquad
 M_-^2  \equiv
\frac{\kappa \X^2}{1 + \kappa \X^2}  M_g^2
\,.
\end{equation} 
In the absence of matter, we find that 
the linear combination of metric perturbations described by 
``$+$'' fields is just the linearized general relativity on 
a de Sitter background, 
while the other linear combination described by ``$-$'' fields
forms a decoupled massive spin-2 field 
with mass $m_{\rm eff}$ around the same background. 

\subsection{Inclusion of matter}

Having finished the analysis on the pure gravity case,
we study the quadratic action taking the matter sector  into account.
Then, in the subsequent subsections 
we study the tensor, vector and scalar sectors in turn, 
and argue the stability conditions for each of them.

We introduce the perturbation of the matter fields as
\begin{equation}
\phi_g = \phi_{g,0}+\delta\phi_g\,,\qquad
\phi_f= \phi_{f,0}+\delta\phi_f\,. 
\end{equation}
We expand the action for the matter fields~(\ref{matter}) up to second order as
\begin{equation}
 I_{{\rm matter},g} = \int d^4x\, N a^3 \sqrt{\thm}\left(L_{{matter},g}^{(0)}+L_{{matter},g}^{(1)}+L_{{matter},g}^{(2)}\right)\,,
\end{equation}
and the second order term is given by 
\begin{eqnarray}
L_{{matter},g}^{(2)} &\! =\! & 
P_g \left(-\frac{\Phi^2}{2}+\frac{1}{2} V^iV_i+\frac{1}{8}\bigl((\Tr
     H)^2-2 H_{ij}H^{ij}\bigr)+\frac{1}{2}\Tr H \Phi\right)
\nonumber \\
&&
+\frac{\rho_g + P_g}{2} \frac{\delta\dot{\phi}_g^2}{\dot{\phi}_{g,0}^2}
+\frac{1-c_g^2}{2 c_g^2}(\rho_g + P_g) \left(\frac{\delta\dot{\phi}_g}{\dot{\phi}_{g,0}}-\Phi\right)^2
+\frac{\rho_g + P_g}{2}
 \left(\Phi-\frac{\delta\dot{\phi}_g}{\dot{\phi}_{g,0}}\right)(2 \Phi-\Tr H)
\nonumber \\
&&
-\frac{N^2(\rho_g + P_g)}{2 a^2 \dot{\phi}_{g,0}^2}\left(D_i
						  \delta\phi_g+\frac{a
\dot{\phi}_{g,0}}{N}V_i\right)\left(D^i \delta\phi_g+\frac{a \dot{\phi}_{g,0}}{N}V^i\right)\,.
\label{action-matterg}
\end{eqnarray}
The action for the $\phi_f$ field can be obtained similarly.

\subsection{Tensor sector}

We start with our analysis on the tensor sector.
We restrict the perturbation of the three metrics 
to the transverse-traceless perturbations as 
\begin{equation}
H_{ij} = H^{TT}_{ij}\,,\qquad h_{ij} = h^{TT}_{ij}\,. 
\end{equation}
To keep the notation simple, we suppress the superscript $TT$ below. 
Combining all terms (two Einstein Hilbert terms, matter terms and the interaction terms), the action quadratic in tensor modes reduces to
\begin{eqnarray}
I^{(2)}_{\rm tensor} &\! =\! & \frac{M_g^2}{8} \int d^4 x \,N a^3\sqrt{\thm} \Bigg[
\frac{\dot{\HTT}^{ij}\dot{\HTT}_{ij}}{N^2}+\frac{\HTT^{ij}}{a^2}\left(\triangle-2K\right)\HTT_{ij}
+
\kappa\, \tildec\, \X^4
\left(\frac{\dot{\hTT}^{ij}\dot{\hTT}_{ij}}{n^2} +\frac{\hTT^{ij}}{\alpha^2}
\left(\triangle-2 K\right)\hTT_{ij}\right) \nonumber\\
&&\qquad\qquad\qquad \qquad\qquad
- m^2 \Gamma 
\left(\HTT^{ij}-\hTT^{ij}\right)\left(\HTT_{ij}-\hTT_{ij}\right)
\Bigg]\,.
\end{eqnarray}
We stress that up to now, we did not assume any branch and only used the
background equations~(\ref{eqh}) and (\ref{eqhf}). In this
most general setup, $\HTT_{ij}$ and $\hTT_{ij}$ fields have a
generically time-dependent coupling. 
Taking the low energy limit where $\X\simeq \X_c$ and $\tildec\simeq 1$, 
the tensor action can be put into the diagonal form
\begin{eqnarray}
I^{(2)}_{\rm tensor} &\! =\! & 
\frac{1}{8}
\int d^4 x \,N a^3\sqrt{\thm}\Bigg[
M_+^2 \left(
\frac{\dot H_+^{ij}\dot H^+_{ij}}{N^2}
+\frac{H_+^{ij}}{a^2} \left(\triangle-2K\right)H^+_{ij}
\right)
\nonumber\\
&&\qquad\qquad\qquad \qquad\qquad 
+M_-^2\left(
\frac{\dot H_-^{ij}\dot H^-_{ij}}{N^2} 
+\frac{H_-^{ij}}{a^2} \left(\triangle-2 K\right)H^-_{ij}
- m_{\rm eff}^2 
H_-^{ij} H^-_{ij}
\right)\Bigg]\,, 
\end{eqnarray}
where $H^\pm_{ij}$ 
are defined as in Eqs.~\eqref{fieldredef}. 
This action is essentially the same as the tensor part 
of Eq.~(\ref{Lpuregrav}) for the pure gravity case.
$H^+_{ij}$ is the massless graviton mode,
and $H^-_{ij}$ is the massive graviton mode 
with mass $m_\text{eff}$ given by Eq.~(\ref{meff}).

\subsection{Vector sector}

We introduce vector perturbations to the metric through
\begin{eqnarray}
&V_i = B_i\,,\qquad
&H_{ij}=\frac{1}{2}\left(D_i E_j+ D_j E_i\right)\,,\nonumber\\
& v_{i} = b_i\,,\qquad
&h_{ij}=\frac{1}{2}\left(D_i S_j + D_j S_i\right)\,,
\end{eqnarray}
where $B_i$, $b_i$, $E_i$ and $S_i$ are transverse with respect to
the covariant differentiation associated with $\thm_{ij}$ metric.
Using the background equations~(\ref{eqn})--(\ref{eqhf}), the total
quadratic action of the vector perturbations becomes
\begin{eqnarray}
I_{\rm vector}^{(2)} &\! =\! & \frac{M_g^2}{8}\int d^4x\,N a^3
 \sqrt{\thm}\Bigg[-\frac{1}{2}\left(\frac{\dot{E}^i}{N}-\frac{2
			       B^i}{a}\right)(\triangle+2
 K)\left(\frac{\dot{E}_i}{N}-\frac{2 B_i}{a}\right)\nonumber\\
&&\qquad\qquad\qquad\qquad\;\;\,
-\frac{\kappa \X^4 \tildec}{2}\left(\frac{\dot{S}^i}{n}-\frac{2 b^i}{\alpha}\right)(\triangle+2K)\left(\frac{\dot{S}_i}{n}-\frac{2b_i}{\alpha}\right)\nonumber\\
&&\qquad\qquad\qquad\qquad\;\;\,
+\frac{m^2 \Gamma }{2} (E^i-S^i)(\triangle+2K)(E_i-S_i)\nonumber\\
&&\qquad\qquad\qquad\qquad\;\;\,+\frac{4 m^2\X J}{\tildec
 +1}(B^i-\tildec b^i)(B_i-\tildec b_i)\Bigg]\,.
\label{actvec}
\end{eqnarray}
In the above form, the action is manifestly gauge invariant, since
$\left(\tfrac{\dot{E}_i}{N}-\tfrac{2 B_i}{a}\right)$,
$\left(\tfrac{\dot{S}_i}{n}-\frac{2 b_i}{\alpha}\right)$, $(E_i-S_i)$,
$(B_i-\tildec b_i)$ are gauge invariant.\footnote{There are actually three independent gauge invariant variables. One can check that the invariance of the first three above implies the invariance of the fourth combination.}

Next, we vary the action with respect to the non-dynamical degrees $B_i$
and $b_i$, to obtain 
\begin{eqnarray}
&&-\frac{4}{a^2} (\triangle+2 K)B_i
 +\frac{2}{a N} (\triangle+2 K) \dot{E}_i+\frac{8m^2 \X J}{\tildec
 +1} (B_i-\tildec \,b_i)=0\,,\nonumber\\
&&-\frac{4}{\alpha^2} (\triangle+2 K)b_i +\frac{2}{\alpha
 n} (\triangle+2 K) \dot{S}_i
-
\frac{8 m^2 J}{\kappa \X^3(\tildec +1)}
(B_i-\tildec\, b_i)=0\,.
\end{eqnarray} %
At this stage, it is convenient to expand the modes in vector harmonics. The solutions to the above equations then read
\begin{equation}
B_i = a\left[\frac{\dot{E}_i}{2 N}
-\frac{{\mathbf A}}{2N}(\dot{E}_i-\dot{S}_i)\right]\,,\qquad
b_i = a\left[\frac{\dot{S}_i}{2 N\tildec}+\frac{{\mathbf
	A}}{2\kappa N\X^2}(\dot{E}_i-\dot{S}_i)\right]\,,
\label{Bbsol}
\end{equation}
where we defined
\begin{equation}
{\mathbf A} \equiv \left[\frac{(k^2-2 K)(\tildec +1)}{2 a^2 m^2 \X
		    J}+\frac{\tildec +\kappa\X^2}{\kappa \X^2}\right]^{-1}\,,
\end{equation}
and $k^2$ is the eigenvalue of $-\Delta$. 

Substituting 
the solution for the auxiliary modes~(\ref{Bbsol}) 
into the action~(\ref{actvec}), we obtain 
\begin{equation}
I_{\rm vector}^{(2)} = \frac{M_-^2}{8}\int dt\,d^3k\,N a^3
{\mathbf A}
\left\{
\frac{\dot{\cal E}^i\dot{\cal E}_i^\star }{N^2}
-\left[ {(k^2-2K)\over a^2 }c_V^2 
   +{\tildec +\kappa\X^2\over 1+\kappa\X^2}m_{\rm eff}^2
\right]
 {\cal E}^i{\cal E}_i^\star\right\}\,,
\end{equation}
where we defined the gauge invariant combination 
\begin{equation}
 {\cal E}_i \equiv 
   (E_i - S_i)\sqrt{
{1+\kappa\X^2\over\kappa\X^2}(k^2-2 K)}
\,, 
\end{equation}
and the squared propagation speed of the vector mode 
\begin{equation}
  c_V^2\equiv {(\tildec +1)\Gamma \over 2\X J}\,. 
\label{cV}
\end{equation}

One can immediately verify that ${\mathbf A}=0$ on the $J=0$ branch 
and thus the vector modes are non-dynamical at linear order.
For the healthy branch, 
$\mathbf A$ should be positive at $k\to \infty$ to 
avoid the ghost instability,
and this condition is satisfied if $J>0$. 
This latter condition is automatically 
satisfied since we have chosen the branch that
satisfies~\eqref{branchcondiJ}. 

If we take the low energy limit, the squared propagation speed 
of perturbation, $c_V^2$, is reduced to 
\begin{equation}
  c_V^2-1\simeq {\tildec -1 \over 2}\left({d\ln J\over d\ln \X}+1 \right)\,. 
\end{equation}
As discussed in Ref.~\cite{DeFelice:2013nba}, 
$d \ln J/d \ln \X$ 
is necessarily large 
for the Vainshtein mechanism to efficiently work in this model.
However, $\tildec -1$ is suppressed at low energies. 
In fact, when $\Gamma$ is dominated by the first term of the right hand
side of (\ref{Gamma}), 
{\it i.e.}, 
when $\Gamma\approx \X J$,  the absolute value of the right hand 
side of the above equation is always less than unity
 and $c_V^2$ is thus guaranteed to be positive. 
 On the other hand, if
we consider the case with 
 $|d \ln J/d \ln \X| > 1$, {\it i.e.}, if the right hand side of
(\ref{Gamma}) is dominated by the second term, then
the right hand side of 
the above expression is $O(m_{\rm eff}^2/m^2 \X J)$, which can be larger 
than $O(1)$. If this is the case, the model parameters are constrained to 
satisfy $d \ln J/d \ln \X > 0$ to avoid the gradient instability.

In the main text, we did not take into account the rotational modes 
of fluids. In Appendix, the case with the matter coupled to the 
$g$-metric only is discussed taking into account the rotational modes. 

\subsection{Scalar Sector}

\subsubsection{Reduction of the quadratic action}
The scalar perturbations are introduced through
\begin{eqnarray}
\Phi\,,\qquad
V_{i}=D_i B\,,\qquad
H_{ij}=2 \thm_{ij} \psi+\left(D_i D_j-\frac{\thm_{ij}}{3} \triangle\right)E\,,\nonumber\\
\varphi\,,\qquad\;\;
v_{i}= D_i b\,,\qquad\;\;
h_{ij}=2 \thm_{ij} \Sigma+\left(D_i D_j-\frac{\thm_{ij}}{3} \triangle\right)S\,,
\end{eqnarray}
in the metrics, and 
\begin{equation}
\phi_g = \phi_{g,0}+\delta\phi_g\,,\qquad
\phi_f= \phi_{f,0}+\delta\phi_f\,,
\end{equation}
in the matter sector. We integrate out the non-dynamical modes $B$, $b$,
$\Phi$ and $\varphi$, ending up with an action depending on six 
variables, $\psi$, $\Sigma$, $E$, $S$, $\delta\phi_g$ and $\delta\phi_f$. 
 As we have not fixed the gauge yet, there are two pure gauge degrees in 
this action. Furthermore, we also expect that 
the would-be Boulware--Deser mode 
 should be non-dynamical. 
Hence three of the six variables are to be eliminated. 

Under the coordinate transformation
\begin{equation}
x^\mu \to x^\mu+\deltaxup^\mu\,,
\end{equation}
where $\deltaxup^\mu = (\deltaxup^0,\nabla^i\deltax)$, the six 
variables transform as
\begin{eqnarray}
\psi\to\psi+N H \deltaxup^0+\frac{1}{3} \triangle \deltax\,,\qquad
E\to E+2 \deltax\,,\qquad
\delta\phi_g\to \delta\phi_g +N \sqrt{X_{g,0}} \deltaxup^0\,,\nonumber\\
\Sigma\to\Sigma+n H_f \deltaxup^0+\frac{1}{3} \triangle \deltax\,,\qquad
S\to S+2 \deltax\,,\qquad
\delta\phi_f\to \delta\phi_f+n \sqrt{X_{f,0}} \deltaxup^0\,,
\label{transformation}
\end{eqnarray}
where
\begin{equation}
X_{g,0} \equiv {\dot{\phi}_{g,0}^2}/{N^2}\,,\qquad
X_{f,0} \equiv {\dot{\phi}_{f,0}^2}/{n^2}\,.
\end{equation}
The three physical degrees of freedom 
can be made manifest by choosing the following gauge invariant variables:
\begin{eqnarray}
Y_1 &\! =\! & \delta\phi_g+\frac{\sqrt{X_{g,0}}}{H} \left(\frac{\triangle}{6} E-\psi\right)\,,\nonumber\\
Y_2 &\! =\! & \delta\phi_f+\frac{\sqrt{X_{f,0}}}{H_f}\left(\frac{\triangle}{6}
						S-\Sigma\right)\,,\nonumber\\
Y_3 &\! =\! & S-E\,. 
\end{eqnarray}
If we adopt the so-called flat gauge conditions $E=\psi=0$, 
which fix the gauge completely as seen from
Eq.~(\ref{transformation}), $Y_1$ and $Y_3$ coincide with
the modes $\delta\phi_g$ and $S$, respectively. In this gauge, 
after expressing $\delta\phi_f$ in terms of $Y_2$, the
Boulware--Deser mode is manifestly non-dynamical 
in the action and can be integrated out. 
The resulting action after expanding the fields
into harmonics is
\begin{equation}
I = \frac{M_g^2}{2}\int dt\,d^3k \,N a^3 \left(\frac{\dot{Y}^\dagger}{N}{\cal K}\frac{\dot{Y}}{N} + \frac{\dot{Y}^\dagger}{N}{\calN}Y-Y^\dagger{\calN}\frac{\dot{Y}}{N} -Y^\dagger\calM Y\right)\,,
\label{actionscalar}
\end{equation}
where $Y = (Y_1, Y_2, Y_3)^T$ is the field array, while ${\cal K}^T =
{\cal K}$, ${\calN}^T = -{\calN}$ and $\calM^T= \calM$ are $3\times3$
real matrices\footnote{
In this paper we do not discuss the effective Newton potential, 
partly because linear analysis is not sufficient to study 
the metric perturbation within the Vainshtein radius. 
Here we quote the result for the Newton potential 
within the quasi-static analysis in the linear perturbation, 
when the matter is coupled only to the 
$g$-metric. 
The resulting Newton potential $\delta\Phi$ is given by 
\begin{equation}
 \delta\Phi=-{\delta\rho_g\over 2M_+^2 (k^2/a^2)}
   \left[{6W+(3+4\kappa\X^2)(k^2/a^2)\over
     6W+3(k^2/a^2)}
     \right]\,.
\end{equation}
}.

\subsubsection{No-ghost conditions}

We first discuss the conditions for avoiding ghost. 
The kinetic matrix can be diagonalized, by applying the rotation
\begin{equation}
R = \left(
\begin{array}{ccccc}
0 && 1 && \dfrac{{\cal K}_{13}{\cal K}_{22}-{\cal K}_{12}{\cal K}_{23}}{{\cal K}_{12}^2-{\cal K}_{11}{\cal K}_{22}}\\\\
1&&-\dfrac{{\cal K}_{12}}{{\cal K}_{22}}&& \dfrac{{\cal K}_{11}{\cal K}_{23}-{\cal K}_{12}{\cal K}_{13}}{{\cal K}_{12}^2-{\cal K}_{11}{\cal K}_{22}}\\\\
0&&0&&1
\end{array}
\right)\,,
\end{equation}
as
\begin{equation}
{\cal K}_{\rm diag} = R^T{\cal K}R = \left(
\begin{array}{lcc}
{\cal K}_{22} & 0&0\\
0& \frac{{\cal K}_{11}{\cal K}_{22}-{\cal K}_{12}^2}{{\cal K}_{22}} &0\\
0 & 0 & \frac{{\rm det}({\cal K})}{{\cal K}_{11}{\cal K}_{22}-{\cal K}_{12}^2}
\end{array}
\right)\,.
\label{Kmatrix}
\end{equation}
Then, we find that the ghost is absent 
if the following inequalities are satisfied 
\begin{equation}
{\rm NG}_1={\cal K}_{22}>0 \,,\qquad
{\rm NG}_2=\frac{{\cal K}_{11}{\cal K}_{22}-{\cal K}_{12}^2}{{\cal K}_{22}}>0\,,\qquad
{\rm NG}_3=\frac{{\rm det}({\cal K})}{{\cal K}_{11}{\cal K}_{22}-{\cal K}_{12}^2}>0\,,
\label{scNG}
\end{equation}
whose explicit expressions are
\begin{eqnarray}
&&{\rm NG}_1 =
\frac{ 2m^2 \kappa H^2 \X J {\cal D}}{\tildec X_{f,0}}
\Biggl[
\frac{2m^2 
\kappa M_g^2 H^2J{\cal D} c_f^2}{\X(\rho_f + P_f)}
+
\biggl( \frac{2K}{a^2}+m^2(\tildec -1)\X J\biggr)
\biggl( \frac{2K}{a^2}-
\frac{m^2\,(\tildec -1)J}{\kappa \X \tildec} 
\biggr)
\nonumber\\
&&\qquad\qquad
+\frac{3K}{k^2-3K}
\biggl\{
\biggl(
\frac{2K}{a^2}-
\frac{m^2\bigl(1+\kappa \X^2\bigr)J}{\kappa \X}
\biggr)^2
- {\cal A} {\cal B}
\biggr\}
+
\frac{9 m^4 M_g^2 J^2 c_g^2}{\kappa\tildec} 
\frac{\bigl({\cal A}-\frac{\tildec k^2}{k^2-3K}{\cal B}\bigr)^2}
{{\cal C}}
\frac{X_{g,0}~{\rm NG}_2}{(\rho_g+P_g)}
\Biggr]^{-1}\,,
\nonumber\\
&&{\rm NG}_2= 
\frac{2(\rho_g+P_g)}{M_g^2 X_{g,0}}\left(
2c_g^2+\frac{(\rho_g+P_g)m^2\X J}{M_g^2 H^2}\frac{\cal D}{\cal C}
\right)^{-1}\,,
\qquad
{\rm NG}_3 =
\frac{3{\cal B}m^2\X J k^2(k^2-3K)}{2{\cal C}}\,,
\label{noghost23}
\end{eqnarray}
where
\begin{eqnarray}
{\cal A} &\equiv& 
\frac{m^2J(\tildec +\kappa \X^2)}{\kappa \X}
-2(\tildec +1)\frac{K}{a^2}
\,,
\qquad
{\cal B} \equiv
\frac{1}{M_g^2(\tildec -1)}\left(\rho_g+P_g-\frac{\X^2}{\kappa}(\rho_f+P_f)\right)\,,\nonumber\\
{\cal C} &\equiv& 
\frac{2k^2}{a^2}\left\{\frac{2(k^2-3K)}{a^2}+3(\tildec +1){\cal B}\right\} + 9{\cal A}{\cal B}\,,
\qquad
{\cal D} \equiv 
\frac{2(\tildec -1)k^2}{a^2}
+3{\cal A}
+\frac{
-3k^2{\cal B}+\frac{K}{m^2\X J}{\cal C}
}{k^2-3K}\,.
\end{eqnarray}
To avoid the catastrophic ghost instability, these conditions 
must be satisfied, at least, in the $k\to\infty$ limit. 
In this limit conditions~(\ref{scNG}) reduce to
\begin{equation}
\text{NG}_1 \to \frac{\X^2 (\rho_f + P_f)}{M_g^2 \tildec c_f^2  X_{f,0}}>0\,,
\qquad
\text{NG}_2 \to \frac{\rho_g + P_g}{M_g^2 c_g^2 X_{g,0}}>0\,,
\qquad
\text{NG}_3 \to \frac38 m^2 a^4 \X J {\cal B} >0\,.
\end{equation}
These conditions are satisfied if 
\begin{equation}
\rho_g + P_g > 0\,,
\qquad
\rho_f + P_f > 0\,,
\qquad
{\cal B}>0\,.
\end{equation}
The first two conditions are just the null energy conditions for the 
matter fields, which are requested for the sound waves of fluids 
to be stable.   

Let us now focus on the last condition, ${\cal B}>0$.  
We can easily verify that ${\cal B}$ can be rewritten as 
\begin{equation}
{\cal B}=2 W+{\X^2\over\kappa M_g^2}\left(\rho_f+P_f \right)\,.
\end{equation}
Therefore, as long as the branch with $W>0$ is concerned, 
the positivity of ${\cal B}$ is guaranteed. 
In the low energy limit or in the pure gravity case, 
we find 
\begin{equation}
{\cal B}
\simeq m_\text{eff}^2 - 2\left(H^2 + \frac{K}{a^2}\right)\,,
\end{equation}
where $m_\text{eff}$ is the effective graviton mass defined in Eq.~(\ref{meff}).
Thus, one can see that the condition ${\cal B}>0$ is the 
extension of the Higuchi bound for the absence of ghost modes.

\subsubsection{Sound speeds}

We now study the speeds of propagation of the scalar modes, 
in the high frequency limit. 
The equation of motion is obtained 
from the action~(\ref{actionscalar}) as
\begin{equation}
{\cal K}\frac{1}{N} \frac{d}{dt}\left(\frac{\dot{Y}}{N}\right) +\left(\frac{\dot{\cal K}}{N}+3H{\cal K}+2{\calN}\right)\frac{\dot{Y}}{N}+\left(\frac{\dot{\calN}}{N}+3H{\calN}+\calM\right)Y=0\,.
\end{equation}
Assuming monochromatic time-dependence of perturbation $Y\propto 
e^{-i\int \omega N dt}$ and the adiabaticity of the background 
$\dot{\omega}/N \ll \omega^2$, 
we obtain the dispersion relation 
\begin{equation}
{\rm det}\left[-\omega^2 {\cal K} - i \omega \left(\frac{\dot{\cal K}}{N}+3H{\cal K}+2{\calN}\right)+\left(\frac{\dot{\calN}}{N}+3H{\calN}+\calM\right)\right]=0\,.
\end{equation}
The 
eigenfrequencies 
can be found by solving this equation. 
Expanding the 
eigenfrequencies 
around $k\to \infty$ as 
$\omega^2 \sim c_s^2 k^2/a^2$, we can read 
the coefficients 
$c_s^2$, the squared propagation speeds 
of perturbation in the high frequency limit, as 
\begin{eqnarray}
c_{s}^2 &\! =\! & 
\frac{m^2}{3{\cal B}}\left(1+
\frac{1}{\kappa \X^2}
\right)\left[4 \Gamma -\X J(\tildec
+1)\right]+\frac{1}{3}\left(\tilde{c}-\frac{4(H^2+\tfrac{K}{a^2})\Gamma }{\X J{\cal B}}\right)\,,\nonumber\\
c_{s\;\rm II}^2 &\! =\! & c_g^2\,,\nonumber\\
c_{s\;\rm III}^2&\! =\! & \tildec^2 c_f^2 \,.
\end{eqnarray}
These must be positive to avoid  a fatal instability.
The positivity of the first one is not obvious. It would be 
instructive 
to rewrite it as 
\begin{eqnarray}
c_{s}^2 -1 &\! =\! &
{2(\tildec -1)\over 3}\frac{d \ln J}{d\ln \X} 
-\frac{2\X^2\left(\rho_f + P_f\right)}{3 \kappa M_g^2 {\cal B}}
\cr
&& +{\tildec -1\over 3{\cal B}}\left[
 2\left(H^2+{K\over a^2}\right)\left({d\ln J\over d\ln \X}-1\right)
 - {\X^2\over \kappa M_g^2}\left(\rho_f + P_f\right)
    \left(2{d\ln J\over d\ln \X}-1\right)
\right]\,. 
\end{eqnarray}
This expression proves that $c_{s}^2 =1$ in the pure gravity 
case. In the low energy limit, the leading order terms come from 
the first line on the right hand side of the above equation and 
they reduce to   
\begin{equation}
c_{s}^2-1 =
{2(\tildec -1)\over 3}\frac{d \ln J}{d\ln \X} 
-\frac{2\X^2\left(\rho_f + P_f\right)}{3 \kappa M_g^2 m_{\rm eff}^2}
+ O\bigl(m^{-4}\bigr)\,.
\end{equation}
Then, the positivity of $c_{s}^2$ 
gives a constraint on the model parameters in a similar sense to $c_V^2$.

Before ending this section, we compare our result with 
 the solutions discussed in \cite{Comelli:2012db}. 
In this case, $\hat \rho_m$ is dominated by the term proportional 
to $1/\xi$, so here (and only here) we assume $\xi \ll 1$. As a result, we have 
$J\sim -\kappa\xi \hat\rho_m=$ constant, hence 
$d \ln J/d \ln \xi \approx 0$. If we assume that there is no matter 
field coupled to the $f$-metric, 
\begin{equation}
 c_s^2\approx -\left(1+2\frac{P_g}{\rho_g}\right), 
\end{equation}
and hence the gradient instability is inevitable. This instability occurs in the regime of validity of the low energy
effective field theory and thus is physical for $m\ll H \ll \Lambda_9$
($\ll \Lambda_3$), where
$\Lambda_n=(M_gm^{n-1})^{1/n}$.~\footnote{To see this, notice that
$\X\simeq \frac{m^2J}{3\kappa H^2}$ and that
$H_f=H/\X\sim H^3/m^2\ll \Lambda_3$ for $H \ll \Lambda_9$,
provided that $\kappa/J=O(1)$. The cutoff scale of the low energy
effective field theory is $\Lambda_3$, provided that $\kappa=O(1)$.}

If a matter sector which couples to the $f$-metric is present and is the dominant contribution of the right hand side in Eq.~(\ref{Fconst1}), a similar calculation yields 
\begin{equation}
 c_s^2\approx -\frac{\left(1+4\frac{P_f}{\rho_f}\right)}{3\left(\frac{P_f}{\rho_f}\right)^2}.
\end{equation}
Again, for both non-relativistic and radiation fluids, the scalar sector suffers from gradient instability.

\section{Summary and Discussion}
\label{sec:summary}
We have presented a linear analysis of cosmological perturbations
in bimetric theory,
in which two metrics are coupled through  
non-derivative coupling so that it does not yield
Boulware-Deser  
ghost as prescribed in~\cite{Hassan:2011zd}.  
We consider perturbations 
around the background of two dynamical FLRW metrics 
sharing spatial isometries, each of which is minimally
coupled to a different k-essence field. 
The contracted
Bianchi identity of either metric yields a constraint which defines two
branches. 
The $J=0$ branch, 
in which the ratio of the scale factors of two 
metrics is fixed to a constant value determined by the condition $J=0$, 
contains only four dynamical degrees of
freedom at linear order~\cite{Comelli:2012db}, while at non-linear
order, is known to suffer from instabilities~\cite{DeFelice:2012mx}. In
this work, we focused on the other, at least, seemingly healthy branch, 
and studied the stability of
the cosmological background against perturbations.

In the absence of matter fields, the linearized action is found to be a
combination of two decoupled spin-2 fields: one of which is linearized
GR, while the other is a linearized spin-2 theory with a mass
term.
In the setup where two matter fields 
are minimally coupled to respective metrics, 
we considered a specific scenario where the mass scale of 
the interaction term between two metrics is large. 
When we choose this mass scale to be small of the order
of the expansion rate $H$, Refs.~\cite{Comelli:2011zm, Comelli:2012db}
concluded that either crossing a singularity or encountering an
instability is inevitable. 
We adopted the basic idea proposed in~\cite{DeFelice:2013nba},  
where 
the singularity and the instability 
are expected to be outside the reach of the low energy effective theory
since the mass parameter is chosen to be large enough. 
Even in such a setup, the extra forces can still be screened under certain
conditions, owing to the Vainshtein mechanism. 
In this paper we showed that
we can actually avoid the singularity and the instability, 
at least, at the level of linear perturbation around FLRW background. 

For the general setup, we identified the necessary
conditions for avoiding ghost instabilities. 
We also derived the expression for the 
propagation speed of each perturbation mode 
and the condition for the avoidance of gradient instabilities. 
We found that the matter perturbations have 
positive kinetic energies as long as the background matter fields do not
violate the null energy condition, 
and that they propagate at the ordinary sound
speeds of the corresponding fluids. 
From the positivity of the 
kinetic term of the gravitational degree of freedom in the scalar
sector, we obtained the extension of
the Higuchi bound~\cite{Higuchi:1986py} (see also \cite{Fasiello:2013woa}), which turns out to be
automatically satisfied as far as the healthy branch is concerned. 
We also found that the squared propagation speeds of gravitational 
degrees of freedom in the scalar and vector sectors can be 
negative, which leads to the so-called gradient instability. 
In order to avoid this instability, 
the model parameters are weakly constrained.

The way how matter couples to gravity is always an issue in
alternative gravity theories. In the present paper we have considered
two matter fields and supposed that each of them minimally couples to
one of the two metrics. This seems a particularly safe choice: in the
massless limit $m\to 0$, the system is decomposed into totally
decoupled two subsystems, each of which consists of a massless
graviton interacting with a matter field. This is trivially consistent
with Weinberg's theorem in 1964~\cite{Weinberg:1964ew} and its
extensions~\cite{Weinberg:1980kq,Porrati:2008rm} that exclude more
than one interacting massless gravitons, since the two subsystems are
completely decoupled from each other in the massless limit. While
other possibilities for matter coupling remain unexplored, any
theoretically consistent schemes of matter coupling should be in
accord with the above mentioned general theorems. For example,
Ref.~\cite{Akrami:2013ffa} proposes that a matter field can couple to
both metrics. In the massless limit of this prescription it appears
that the two massless gravitons can interact with each other through
the matter field. It is interesting to see whether and how this can be
reconciled with the general theorems.  

Finally, the present study should help generalizing the predictions of the bimetric theory on future experiments (see {\it e.g.}\
\cite{Konnig:2014dna}).

{\bf Note added:} Shortly after this work was completed, Ref.~\cite{Solomon:2014dua} appeared on the arXiv, where solutions with effective mass of the order of the present-day Hubble rate are considered (see also \cite{vonStrauss:2011mq,Berg:2012kn,Akrami:2012vf,Konnig:2013gxa}). We remark that our analysis does not exclude healthy backgrounds other than the ones considered in the present work.

\begin{acknowledgments}
\quad
A.E.G. thanks Marco Crisostomi for useful discussions.
N.T.\ thanks Kazuya Koyama and Yi Wang for useful discussions.
A.E.G.\ acknowledges financial support from the European Research
Council under the European Union's Seventh Framework Programme
(FP7/2007-2013) / ERC Grant Agreement n. 306425 ``Challenging General
Relativity''.  The work of A.E.G., S.M., N.T.\ was supported by the
World Premier International Research Center Initiative (WPI
Initiative), MEXT, Japan. N.T.\ acknowledges JSPS Grant-in-Aid for
Scientific Research 25$\cdot$755. This work is in part supported by
the Grant-in-Aid for the Global COE Program ``The Next Generation of
Physics, Spun from Universality and Emergence'' from the Ministry of
Education, Culture, Sports, Science and Technology (MEXT) of Japan and
also by Grant-in-Aid for Scientific Research Nos.~24540256, 24103006,
24103001, 21244033 and 21111006.  Part of this work took place during
the workshop YITP-X-13-03.

\end{acknowledgments}

\appendix

\section{Vector modes}

In this appendix we discuss more in detail the contribution of
the physical perfect fluid to the vector modes. Here, we will
only consider the simple case with $K=0$ in the absence of the hidden
matter field $\phi_f$. Instead, we replace the scalar
fluid $\phi_g$ with a standard perfect fluid. 
In order to achieve this last step we
follow the procedure by Schutz outlined 
in~\cite{Schutz:1970my}. On setting the gauge $E_i=0$, the physical
metric perturbation variables are $\delta g_{0i}=N a V_i$. As for the
second metric, instead, $\delta f_{0i}=n\alpha v_i$, and 
$\delta f_{ij}=\frac12 \tilde a^2(S_{i,j}+S_{j,i})$, 
and we introduce the
gauge invariant variable $\tilde
V_i=-\frac{N\alpha^2}{2an}(\dot S_i/n-2v_i/\alpha)$. Finally, the
perfect fluid has the velocity perturbation $\delta u_i$. 
All these variables satisfy the usual transverse condition.

On using the Schutz' Lagrangian for the perfect fluids\footnote{We model here 
$u_\alpha=\mu^{-1}(\partial_\alpha\ell+A_1\partial_\alpha
B_1+A_2\partial_\alpha B_2)$ for a perfect fluid with equations of state
$p=w\rho$, and $\mu=\mu_0 (n/n_0)^w$. We can then define the gauge
invariant combination $F_i\equiv
\frac{E_{i}}2-\frac{b_{1i}}{\vec{b}_{1}\cdot\vec{b}_{1}} \delta 
B_{1}-\frac{b_{2i}}{\vec{b}_{2}\cdot\vec{b}_{2}} \delta B_{2}$, with
$\vec b_1\cdot\vec b_2=0$, $\vec b_1\cdot\vec k=0=\vec b_2\cdot \vec k$,
and, on the background $A_1=0=A_2$, together with $B_{1,i}=b_{1i}$,
$B_{2,i}=b_{2i}$.}, 
we find that the action can be written as 
\begin{eqnarray}
I&\! =\! &\int N dt d^3x\left\{ \frac{aM_g^2}4
 (\partial_j V_i)(\partial_j V_i)-\frac12 (\rho+p) a \delta u_i \delta u_i
+(\rho+p) a^2\delta u_i(a\dot F_i/N+V_i)\right.\nonumber\\
&&{}+\frac{aM_f^2\X^2r^3}4(\partial_j \tilde V_i)(\partial_j \tilde V_i)
+\frac{m^2M_g^2J\X a^3}{2(\tildec +1)} V_iV_i
+\frac{r^4a^3m^2M_g^2J\X}{2(\tildec +1)} \tilde V_i\tilde V_i\nonumber\\
&&{}-V_i\left[\frac{a^4J\X M_g^2 m^2}{2N(\tildec +1)}\dot{S}_i
+\frac{a^3J\X m^2 M_g^2r^2}{\tildec +1} \tilde V_i\right]\nonumber\\
&&{}+ \frac{a^4J\X M_g^2 m^2r^2}{2N(\tildec +1)} \tilde V_i \dot{S}_i
+\frac{a^5 J\X m^2 M_g^2}{8N^2(\tildec +1)} \dot{S}_i\dot{S}_i\nonumber\\
&&{}+\left.\frac{a^3}4\left[M_f^2\X^2(\tildec -1) H^2+\frac{M_f^2\X^2}2\frac{\rho+p}{M_g^2}
-{1\over 8}J\X M_g^2 m^2 
\left[2+(\tildec -1)\left(4+\frac{d\ln J}{d\ln \X}+4\frac{M_f^2}{M_g^2}
		     \X^2\right)\right]\right](\partial_j
S_i)(\partial_j S_i)\right\}\,.\cr
&&
\label{eq:azVec}
\end{eqnarray}
As for the perfect fluid, we can find that the equation of motion for 
$F_i$ leads to
\begin{equation}
(\rho+p)a^3\delta u_i=\textrm{constant}\,.
\end{equation}
The first line of Eq.~(\ref{eq:azVec}) corresponds to the General
Relativity result. Anything else comes from the bimetric theory. By
removing the auxiliary fields $\delta u_i$, $V_i$, and $\tilde V_i$
(by using Fourier decomposition) we find the reduced action for four
independent modes (equivalent two by two) as 
\begin{equation}
I=\int Ndt d^3 x\left[A_{11}\dot F_i\dot F_i+A_{22}\dot{S}_i\dot{S}_i
+2A_{12}\dot{F}_i\dot{S}_i
-E_{22}S_iS_i\right]\,.
\end{equation}
For high $k$'s the no-ghost conditions read
\begin{equation}
A_{22}\approx\frac{m^2 M_g^2J\X a^5}{8N^2(\tildec +1)}>0\,,\qquad
A_{11}A_{22}-A_{12}^2\approx\frac{a^{10}m^2J\X M_g^2(\rho+p)}{16N^4(\tildec +1)}>0\,.
\end{equation}
On the chosen background these conditions imply $J>0$. 
This condition is automatically satisfied by the healthy branch
background solution. 

We find two different speeds of propagation. Namely, one is 
$c^2_{V}$ in Eq.~\eqref{cV} and the other is 
\begin{eqnarray}
c^2_{V,2}&\! =\! &0\,.
\end{eqnarray}
The second mode corresponds to the degree of freedom in the matter
sector.

\end{document}